\documentclass[prd,showpacs,showkeys,nofootinbib,twocolumn]{revtex4-1}
\usepackage{amsmath,amsfonts,amssymb}
\usepackage{xcolor}
\usepackage[colorlinks=true,urlcolor=blue,linkcolor=blue,citecolor=blue]{hyperref}

\begin{document}

\title{Demystifying autoparallels in alternative gravity}

\author{Yuri N. Obukhov}
\email{obukhov@ibrae.ac.ru}
\affiliation{Theoretical Physics Laboratory, Nuclear Safety Institute, 
Russian Academy of Sciences, B.Tulskaya 52, 115191 Moscow, Russia} 

\author{Dirk Puetzfeld}
\email{dirk.puetzfeld@zarm.uni-bremen.de}
\homepage{http://puetzfeld.org}
\affiliation{ZARM, University of Bremen, Am Fallturm, 28359 Bremen, Germany} 

\date{ \today}

\begin{abstract}
Autoparallel curves along with geodesic curves can arise as trajectories of physical test bodies. We explicitly derive autoparallels as effective post-Riemannian geometric constructs, and at the same time we argue \emph{against} postulating autoparallels as fundamental equations of motion for test bodies in alternative gravity theories. 
\end{abstract}


\maketitle


\section{Introduction}\label{introduction_sec}

The equations of motion of test bodies in relativistic gravity are tightly linked to the conservation laws of the theory \cite{Mathisson:1937,Papapetrou:1951:3,Dixon:2015}. The explicit derivation of these equations has been intertwined with the development of approximation schemes within relativistic gravity \cite{Ruse:1930:1,Synge:1960,DeWitt:Brehme:1960}. As it is well known, geodesic curves arise as trajectories of structureless test bodies in Riemannian spacetimes with the metric $g_{ij}$ as the gravitational field potential, that determines the metric-compatible Christoffel connection $\widetilde{\Gamma}{}_{ij}{}^k = {\frac 12}g^{kl}\left(\partial_i g_{jl} + \partial_j g_{il} - \partial_l g_{ij}\right)$.

In alternative gravity theories the set of the gravitational field variables is extended to ($g_{ij}$, $\Gamma_{ij}{}^k$) where the connection and the metric are no longer compatible, so that the torsion $T_{ij}{}^k := \Gamma_{ij}{}^k - \Gamma_{ji}{}^k$ and the nonmetricity $Q_{kij} := -\,\nabla_k g_{ij} = -\,\partial_k g_{ij} + \Gamma_{ki}{}^l g_{lj} + \Gamma_{kj}{}^l g_{il}$ are nontrivial, in general. Autoparallel curves have been \emph{postulated} on several occasions in the literature as candidates for the equations of motion of test bodies in alternative gravity theories. Such ad-hoc postulates, unsubstantiated by the conservation laws, usually lead to inconsistencies with the field equations. Consequently one should abstain from the practice of postulating equations of motion instead of deriving them \cite{Obukhov:Puetzfeld:2015:1,Jimenez:Delholm:2020:1}. With this warning in mind, we here report on two \emph{special} cases, in which autoparallel curves actually do emerge in theories with post-Riemannian spacetime structure.

Let us consider the dynamics of massive particles under the action of the gravitational $g_{ij}$ and a scalar $\phi$ field. We demonstrate that it is possible to recast the latter into a geometric property of the underlying spacetime, and construct an effective torsion $T_{ij}{}^k$ and nonmetricity $Q_{kij}$ from this scalar field. As a preliminary step, we recall that the deviation of the spacetime geometry from the Riemannian \cite{Schouten:1954} one is described by the distortion tensor which measures the difference of connections, $N_{ij}{}^k := \widetilde{\Gamma}{}_{ij}{}^k - \Gamma_{ij}{}^k$. Explicitly:
\begin{eqnarray}
N_{ij}{}^k &=&- \, {\frac 12}\left(T_{ij}{}^k - T_j{}^k{}_i + T^k{}_{ij}\right) \nonumber \\
&&+ {\frac 12} \left(Q^k{}_{ij} - Q_{ij}{}^k - Q_{ji}{}^k\right).\label{N}
\end{eqnarray}

\section{Effective torsion from the scalar field}

At first, we set the nonmetricity equal zero $Q_{kij} = 0$, and introduce the torsion tensor of the form
\begin{equation}
T_{ij}{}^k = \delta_i^k V_j - \delta_j^k V_i,\label{TV}
\end{equation}
where the vector field 
\begin{equation}
V_i = \xi\partial_i\phi\label{Vi}
\end{equation} 
is the gradient of the scalar field $\phi$ with some arbitrary parameter $\xi$. By construction, the torsion (\ref{TV}) has only one irreducible part, namely, the trace $T_{ki}{}^k = 3V_i$. 

Accordingly, we derive from (\ref{TV}) and (\ref{N}) the distortion (which is equal to the contortion tensor in this case):
\begin{equation}
N_{ij}{}^k =  g_{ij} V^k - \delta_i^k V_j.\label{KV}
\end{equation}
Now we write down the autoparallel equation for a point particle with the velocity $u^i = {\frac {dx^i}{ds}}$:
\begin{equation}
{\frac {Du^k}{ds}} = {\frac {d^2x^k}{ds^2}} + \Gamma_{ij}{}^ku^iu^j = 0.\label{auto}
\end{equation}
Substituting $\Gamma_{ij}{}^k = \widetilde{\Gamma}{}_{ij}{}^k - N_{ij}{}^k$ and using (\ref{KV}), we recast (\ref{auto}) into
\begin{equation}
{\frac {d^2x^k}{ds^2}} + \widetilde{\Gamma}_{ij}{}^ku^iu^j - \left(V^k - u^ku^iV_i\right)= 0.\label{e1}
\end{equation}
With the help of (\ref{Vi}), we finally rewrite (\ref{e1}) as 
\begin{equation}
{\frac {d^2x^k}{ds^2}} + \widetilde{\Gamma}_{ij}{}^ku^iu^j = \xi\left(g^{ik} - u^ku^i\right)\partial_i\phi.\label{e2}
\end{equation}

\section{Effective nonmetricity from the scalar field}

Another non-Riemannian interpretation of the equation of motion (\ref{e2}) can be achieved by setting the torsion equal zero $T_{ij}{}^k = 0$, and by using the nonmetricity of the form
\begin{equation}
Q_{kij} = -\,g_{ij}V_k + g_{k(i}V_{j)}.\label{QV}
\end{equation}
As becomes clear from (\ref{QV}), such nonmetricity has two nontrivial irreducible parts: the 3rd and the 4th (Weyl vector). Substituting this into (\ref{N}), we find the distortion tensor
\begin{equation}
N_{ij}{}^k = g_{ij} V^k - \delta_{(i}^k V_{j)}.\label{NQ}
\end{equation}
As a result, we discover that the autoparallel equation (\ref{auto}) again reads
\begin{equation}
{\frac {d^2x^k}{ds^2}} + \widetilde{\Gamma}_{ij}{}^ku^iu^j 
- \left(V^k - u^ku^iV_i\right)= 0.\label{e4}
\end{equation}
Then, by identifying, as before, the vector field with the gradient of the scalar field (\ref{Vi}), we finally obtain the same equation of motion (\ref{e2}).

\section{More on autoparallels}

The possible role of autoparallels in alternative gravity has been the subject to several works \cite{Hehl:1971,Ponomarev:1974,Adamowicz:Trautman:1975,vonderHeyde:1975}. As becomes clear from (\ref{N}), the most general autoparallel equation may be written in terms of the metric, the nonmetricity, and the torsion and the well-known decompositions \cite{Schouten:1954,Hehl:1995} of these geometrical quantities may be used to study subcases.
\begin{eqnarray}
&&{\frac {d^2x^k}{ds^2}} + \left(\widetilde{\Gamma}{}_{ij}{}^k - N_{ij}{}^k \right) u^iu^j = 0 \label{auto_decomp}
\end{eqnarray}
In view of the symmetry of the factor $u^iu^j$, all four irreducible parts of the nonmetricity tensor $Q_{ijk}={}^{(1)}Q_{ijk}+{}^{(2)}Q_{ijk}+{}^{(3)}Q_{ijk}+{}^{(4)}Q_{ijk}$ and two irreducible pieces of the torsion tensor $T_{ijk}={}^{(1)}T_{ijk}+{}^{(2)}T_{ijk}$ (the traceless tensor part, and the vector trace of the torsion) can contribute to (\ref{auto_decomp}), excluding the axial trace of the torsion. It is possible to develop a corresponding classification of the autoparallel curves, thereby generalizing the earlier findings of \cite{Capozziello:etal:2001} from Riemann-Cartan spaces to metric-affine geometries. Here we have demonstrated the distinguished role of the vector parts of the torsion and the nonmetricity.

However, it should be stressed, as it has been shown for a very large class of gravitational theories \cite{Puetzfeld:Obukhov:2014}, that the autoparallel equation does {\it not} emerge as the equation of motion.

One should carefully distinguish the physical and the mathematical aspects when analyzing the motion of test particles. From the mathematical point of view, it is possible to postulate any trajectories in the post-Riemannian spacetimes, and autoparallel curves may look indeed like a reasonable choice which describes the ``straightest'' curve between any two points on a manifold. However, from the point of view of physics, we should explain the physical reason which forces a test particle to move along such a trajectory. In this sense, the following analogy could be helpful. Let us consider an electrically neutral but massive point particle on a Riemannian manifold, which moves, as one knows, along the Riemannian geodesic curves. This motion is a result of an action of the gravitational field on particle's mass. Suppose, however, that in addition we switch an electric or magnetic (or, in general, an electromagnetic) field on. Does this change the motion of a neutral test particle? The answer is crystal clear: no. Such a particle will still move along the Riemannian geodesic trajectory, without feeling the presence of electric and magnetic fields, no matter how strong they are, because it does not carry an electric charge. The situation with the motion of a massive point particle in post-Riemannian geometries is completely analogous. Since the mass is the only physical property (= ``gravitational charge'') of a particle, the latter plainly cannot feel any additional post-Riemannian geometrical fields, such as the torsion and the nonmetricity. Just like one needs to attach an electric charge to a test particle in order to force it to deviate from the geodetic motion under the action of an electromagnetic field, in order to force a particle to move along an autoparallel instead of a geodesic curve, one should attach to it an additional ``charge'' that interacts with the torsion and/or nonmetricity. Indeed, the corresponding analysis \cite{Puetzfeld:Obukhov:2014,Obukhov:Puetzfeld:2015:1} of the equations of motion in metric-affine gravity (MAG) reveals that one needs matter with microstructure (i.e., with spin, dilation and/or shear charge) to feel post-Riemannian parts of the gravitational field. 

\section{Conclusion}

In order to demonstrate the physical relevance of the equation of motion above, let us consider the modified action of a structureless point particle
\begin{equation}
I = \int \varphi\,ds = \int \varphi(x)\,\sqrt{g_{ij}(x){\frac {dx^i}{d\lambda}} {\frac {dx^j}{d\lambda}}}\,d\lambda.\label{Ids}
\end{equation}
Such a nonminimal coupling, where the function $\varphi$ can depend arbitrary on the spacetime coordinates (either directly, or via the geometric invariants), can be viewed as a description of the motion of a body with a variable mass \cite{Brans:Dicke:1961}. 

The variation with respect to the coordinates $x^k(s)$ of the particle is straightforwardly computed
\begin{eqnarray}
&&\delta I = \int ds\left[\varphi\left({\frac 12}\delta g_{ij}{\frac {dx^i}{ds}} {\frac {dx^j}{ds}} + g_{ij}{\frac {dx^i}{ds}}{\frac {d\delta x^j}{ds}}\right) + \delta\varphi\right]\nonumber\\
&&= \int ds\,\delta x^k\Bigg[{\frac 12}\varphi(\partial_kg_{ij})u^iu^j - {\frac d{ds}} (\varphi g_{ki})u^i - \varphi g_{ki}{\frac {du^i}{ds}} + \partial_k\varphi\Bigg]\nonumber\\
&&= -\,\int ds\,g_{ik}\delta x^k\Bigg[\varphi\left({\frac {du^i}{ds}} + \widetilde{\Gamma}_{mn}{}^iu^mu^n\right) - \partial^i\varphi + {\frac {d\varphi}{ds}}u^i\Bigg]. \nonumber \\ \label{dI}
\end{eqnarray}

Accordingly, the minimal action principle $\delta I = 0$ yields the equation of motion
\begin{equation}
{\frac {d^2x^i}{ds^2}} + \widetilde{\Gamma}_{mn}{}^iu^mu^n = \left(g^{ij} - u^iu^j\right) \partial_j\log\varphi. \label{e3}
\end{equation}

By setting $\varphi = e^{\xi\phi}$, we thus conclude that {\it formally} there exist two possibilities to recast the equation of motion (\ref{e3}) into an \emph{autoparallel} in a non-Riemannian spacetime: with the effective torsion (\ref{TV}) or with the effective nonmetricity (\ref{QV}). In principle, one could even think of a combination of the two cases into a more general spacetime geometry with both torsion and nonmetricity.

It is instructive to compare the above results to the equations of motion of a spinless particle nonminimally coupled to the gravitational field, derived from first principles on the basis of the conservation laws \cite{Puetzfeld:Obukhov:2013:1,Puetzfeld:Obukhov:2013:2,Hehl:Obukhov:Puetzfeld:2013:1}. Therein, by means of multipolar techniques it was shown, for a very large class of gravitational theories, that test bodies perform a non-geodesic motion with a ``pressure''-like force. At the first sight, the resulting equations are equivalent to the autoparallel curve (\ref{auto}) found above. However, one should be clear about the fact that this similarity is \emph{purely formal}. In contrast to the derivation given in \cite{Puetzfeld:Obukhov:2013:1,Puetzfeld:Obukhov:2013:2,Hehl:Obukhov:Puetzfeld:2013:1}, the effective torsion (\ref{TV}) and the nonmetricity (\ref{QV}) are \emph{not} linked to any kind of gravitational field equations. 

Alternative theories of gravity can be studied in the unified framework of metric-affine gravity \cite{Hehl:1995}. The latter is based on gauge-theoretic principles, and it takes into account microstructural properties of matter (spin, dilation current, proper hypercharge) as possible physical sources of the gravitational field on an equal footing with macroscopic properties (energy and momentum) of matter. The corresponding spacetime landscape includes as special cases the geometries of Riemann, Riemann-Cartan, Weyl, and Weitzenb\"ock. In the standard formulation of MAG as a gauge theory \cite{Hehl:1995}, the gravitational gauge potentials are identified with the metric, coframe, and the linear connection. The corresponding gravitational field strengths are then the nonmetricity, the torsion, and the curvature, respectively. A unified covariant framework for the test body equations of motion in MAG can be found in \cite{Puetzfeld:Obukhov:2014}. 

As long as no dynamics of the genuine spacetime torsion and nonmetricity is assumed in a particular alternative gravity model, it makes \emph{no} sense to postulate the equation of motion of a test body in the form of an autoparallel. Such a postulate may even be at odds with the actual equations of motion of a theory -- which should be derived from its conservation laws, and for the minimal coupling of the structureless matter the resulting trajectory is a Riemannian geodesic curve \cite{Puetzfeld:Obukhov:2014,Obukhov:Puetzfeld:2015:1}. 

It would be a deep delusion to overestimate the two curious examples with artificial torsion and nonmetricity above, and interpret them as an evidence of the physical significance of the autoparallel prescription.
 
\begin{acknowledgments}
This work was supported by the Deutsche Forsch\-ungsgemeinschaft (DFG) through the grant PU 461/1-2 – project number 369402949 (D.P.). The work of Y.N.O. was partially supported by the Russian Foundation for Basic Research (Grant No. 18-02-40056-mega).
\end{acknowledgments}

\bibliographystyle{unsrtnat}
\bibliography{demystauto}

\begin{thebibliography}{20}
\providecommand{\natexlab}[1]{#1}
\providecommand{\url}[1]{\texttt{#1}}
\expandafter\ifx\csname urlstyle\endcsname\relax
  \providecommand{\doi}[1]{doi: #1}\else
  \providecommand{\doi}{doi: \begingroup \urlstyle{rm}\Url}\fi

\bibitem[Mathisson(1937)]{Mathisson:1937}
M.~Mathisson.
\newblock {Neue Mechanik materieller Systeme}.
\newblock \emph{Acta Phys. Pol.}, 6:\penalty0 163, 1937.
\newblock URL \url{https://doi.org/10.1007/s10714-010-0939-y}.

\bibitem[Papapetrou(1951)]{Papapetrou:1951:3}
A.~Papapetrou.
\newblock {Spinning test-particles in General Relativity. I}.
\newblock \emph{Proc. Roy. Soc. Lond. A}, 209:\penalty0 248, 1951.
\newblock URL \url{https://doi.org/10.1098/rspa.1951.0200}.

\bibitem[{Dixon}(2015)]{Dixon:2015}
W.~G. {Dixon}.
\newblock {The New Mechanics of Myron Mathisson and its subsequent
  development}.
\newblock \emph{''Equations of Motion in Relativistic Gravity'', D. Puetzfeld
  et. al. (eds.), Fundamental theories of Physics, Springer}, 179:\penalty0 1,
  2015.
\newblock URL \url{https://doi.org/10.1007/978-3-319-18335-0_1}.

\bibitem[{Ruse}(1930)]{Ruse:1930:1}
H.~S. {Ruse}.
\newblock {Taylor's theorem in the tensor calculus}.
\newblock \emph{Proc. Lond. Math. Soc.}, 31:\penalty0 225, 1930.
\newblock URL \url{https://doi.org/10.1112/plms/s2-32.1.87}.

\bibitem[{Synge}(1960)]{Synge:1960}
J.~L. {Synge}.
\newblock \emph{{Relativity: The general theory}}.
\newblock North-Holland, Amsterdam, 1960.

\bibitem[{DeWitt} and {Brehme}(1960)]{DeWitt:Brehme:1960}
B.~S. {DeWitt} and R.~W. {Brehme}.
\newblock {Radiation damping in a gravitational field}.
\newblock \emph{Ann. Phys (NY)}, 9:\penalty0 220, 1960.
\newblock URL \url{https://doi.org/10.1016/0003-4916(60)90030-0}.

\bibitem[{Obukhov} and {Puetzfeld}(2015)]{Obukhov:Puetzfeld:2015:1}
Yu.~N. {Obukhov} and D.~{Puetzfeld}.
\newblock {Multipolar test body equations of motion in generalized gravity
  theories}.
\newblock \emph{''Equations of Motion in Relativistic Gravity'', D. Puetzfeld
  et. al. (eds.), Fundamental theories of Physics, Springer}, 179:\penalty0 67,
  2015.
\newblock URL \url{https://doi.org/10.1007/978-3-319-18335-0_2}.

\bibitem[{Jim\'enez} and {Delhom}(2020)]{Jimenez:Delholm:2020:1}
J.~B. {Jim\'enez} and A.~{Delhom}.
\newblock {Instabilities in metric-affine theories of gravity with higher order
  curvature terms}.
\newblock \emph{Eur. Phys. J. C}, 80:\penalty0 585, 2020.
\newblock URL \url{https://doi.org/10.1140/epjc/s10052-020-8143-z}.

\bibitem[{Schouten}(1954)]{Schouten:1954}
J.~A. {Schouten}.
\newblock \emph{{Ricci-Calculus. An introduction to tensor analysis and its
  geometric applications}}.
\newblock Springer, Berlin, 2nd edition, 1954.
\newblock URL \url{https://doi.org/10.1007/978-3-662-12927-2}.

\bibitem[{Hehl}(1971)]{Hehl:1971}
F.~W. {Hehl}.
\newblock {How does one measure torsion of space-time?}
\newblock \emph{Phys. Lett. A}, 36:\penalty0 225, 1971.
\newblock URL \url{https://doi.org/10.1016/0375-9601(71)90433-6}.

\bibitem[{Ponomariev}(1974)]{Ponomarev:1974}
V.N. {Ponomariev}.
\newblock {Observable effects of space-time torsion}.
\newblock \emph{Moscow Univ. Phys. Bull.}, 29\penalty0 (5):\penalty0 139, 1974.

\bibitem[{Adamowicz} and {Trautman}(1975)]{Adamowicz:Trautman:1975}
W.~{Adamowicz} and A.~{Trautman}.
\newblock {The principle of equivalence for spin}.
\newblock \emph{Bull. Acad. Pol. Sci.}, 23:\penalty0 339, 1975.

\bibitem[{von der Heyde}(1975)]{vonderHeyde:1975}
P.~{von der Heyde}.
\newblock {The equivalence principle in the $U_4$ theory of gravitation}.
\newblock \emph{Lett. Nuov. Cim.}, 14:\penalty0 250, 1975.
\newblock URL \url{https://doi.org/10.1007/BF02745635}.

\bibitem[{Hehl} et~al.(1995){Hehl}, {McCrea}, {Mielke}, and
  {Ne'eman}]{Hehl:1995}
F.~W. {Hehl}, J.~D. {McCrea}, E.~W. {Mielke}, and Y.~{Ne'eman}.
\newblock {Metric-affine gauge theory of gravity: Field equations, Noether
  identities, world spinors, and breaking of dilation invariance}.
\newblock \emph{Phys. Rep.}, 258:\penalty0 1, 1995.
\newblock URL \url{https://doi.org/10.1016/0370-1573(94)00111-F}.

\bibitem[{Capozziello} et~al.(2001){Capozziello}, {Lambiase}, and
  {Stornaiolo}]{Capozziello:etal:2001}
S.~{Capozziello}, G.~{Lambiase}, and C.~{Stornaiolo}.
\newblock {Geometric classification of the torsion tensor of space-time}.
\newblock \emph{Ann. Phys. (Berlin)}, 10:\penalty0 713, 2001.
\newblock URL
  \url{https://doi.org/10.1002/1521-3889(200108)10:8<713::AID-ANDP713>3.0.CO;2-2}.

\bibitem[{Puetzfeld} and {Obukhov}(2014)]{Puetzfeld:Obukhov:2014}
D.~{Puetzfeld} and Yu.~N. {Obukhov}.
\newblock {Equations of motion in metric-affine gravity: a covariant unified
  framework}.
\newblock \emph{Phys. Rev. D}, 90:\penalty0 084034, 2014.
\newblock URL \url{https://doi.org/10.1103/PhysRevD.90.084034}.

\bibitem[{Brans} and {Dicke}(1961)]{Brans:Dicke:1961}
C.~{Brans} and R.~H. {Dicke}.
\newblock {Mach's principle and a relativistic theory of gravitation}.
\newblock \emph{Phys. Rev.}, 124:\penalty0 925, 1961.
\newblock URL \url{https://doi.org/10.1103/PhysRev.124.925}.

\bibitem[{Puetzfeld} and
  {Obukhov}(2013{\natexlab{a}})]{Puetzfeld:Obukhov:2013:1}
D.~{Puetzfeld} and Yu.~N. {Obukhov}.
\newblock {Equations of motion in gravity theories with nonminimal coupling: a
  loophole to detect torsion macroscopically?}
\newblock \emph{Phys. Rev. D}, 88:\penalty0 064025, 2013{\natexlab{a}}.
\newblock URL \url{https://dx.doi.org/10.1103/PhysRevD.88.064025}.

\bibitem[{Puetzfeld} and
  {Obukhov}(2013{\natexlab{b}})]{Puetzfeld:Obukhov:2013:2}
D.~{Puetzfeld} and Yu.~N. {Obukhov}.
\newblock {Unraveling gravity beyond Einstein with extended test bodies}.
\newblock \emph{Phys. Lett. A}, 377:\penalty0 2447, 2013{\natexlab{b}}.
\newblock URL \url{https://dx.doi.org/10.1016/j.physleta.2013.07.024}.

\bibitem[{Hehl} et~al.(2013){Hehl}, {Obukhov}, and
  {Puetzfeld}]{Hehl:Obukhov:Puetzfeld:2013:1}
F.~W. {Hehl}, Yu.~N. {Obukhov}, and D.~{Puetzfeld}.
\newblock {On Poincar\'{e} gauge theory of gravity, its equations of motion,
  and Gravity Probe B}.
\newblock \emph{Phys. Lett. A}, 377:\penalty0 1775, 2013.
\newblock URL \url{https://dx.doi.org/10.1016/j.physleta.2013.04.055}.

\end{thebibliography}
\end{document}